\documentclass[aps,prc,tightenlines,showpacs,preprint,
floatfix,amssymb,byrevtex,nofootinbib]{revtex4}  

\usepackage{epsfig,bm,dcolumn}

\begin{document}

\preprint{hep-ph/0508251}


\title{$\bm{D_s (2317)}$ as a four-quark state in QCD sum rules}


\author{Hungchong Kim}
\email{hungchon@postech.ac.kr}

\affiliation{Department of Physics, Pohang University of Science and
Technology, Pohang 790-784, Korea}

\author{Yongseok Oh}
\email{yoh@physast.uga.edu}

\affiliation{Department of Physics and Astronomy,
University of Georgia, Athens, Georgia 30602, U.S.A.}



\begin{abstract}

We perform a QCD sum rule study of the open-charmed $D_s (2317)$
as a four-quark state.
Using the diquark-antidiquark picture for the four-quark state, we
consider four possible interpolating fields for $D_s (2317)$, namely,
scalar-scalar, pseudoscalar-pseudoscalar, vector-vector, and 
axial-vector--axial-vector types.
We test all four currents by constructing four separate sum rules.
The sum rule with the scalar-scalar current gives a stable value for
the $D_s$ mass which qualitatively agrees with the experimental value,
and the result is not sensitive to the continuum threshold.
The vector-vector sum rule also gives a stable result with small sensitivity
to the continuum threshold and the extracted mass is somewhat lower than
the scalar-scalar current value. 
On the other hand, the two sum rules in the pseudoscalar and axial-vector
channels are found to yield the mass highly sensitive to the continuum
threshold, which implies that a four-quark state with the combination of
pseudoscalar-pseudoscalar or axial-vector--axial-vector type would be
disfavored.
These results would indicate that $D_s (2317)$ is a bound state of
scalar-diquark and scalar-antidiquark and/or vector-diquark and
vector-antidiquark.

\end{abstract}

\pacs{14.40.Lb, 11.55.Hx, 12.38.Lg}

\maketitle

\section{Introduction}

Recently new resonances with open- and hidden-charmed states have been
reported by various experiments.
The reported open-charmed states include $D (2308)$~\cite{Abe:2003zm},
$D_s (2317)$~\cite{Aubert:2003fg},
$D_s (2460)$~\cite{Besson:2003cp,Abe:2003jk}, and
$D_s (2632)$~\cite{Evdokimov:2004iy},
while $X (3872)$~\cite{Choi:2003ue} and $Y (4260)$~\cite{Aubert:2005rm}
have been reported as possible hidden-charmed states.
This series of new resonances in charm sector opens a new challenge for
heavy quark system and a systematic analysis is required to understand
their internal structure and their properties.

Among the open-charmed states, $D_s (2317)$, which is believed to have
the quantum number $J^P=0^+$~\cite{PDG04}, is particularly interesting.
It might be interpreted as a two-quark state of
$c\bar{s}$~\cite{BEH03,NRZ03,DHLZ03,LLMP04},
a four-quark state~\cite{CH03,BPP03,Dmit04-05,BLMNN:05},
a molecular state of $DK$~\cite{CL04}, or a $D_s (1968) \pi$
atom~\cite{Szczepaniak03}.
The exotic possibility like a four-quark state is quite intriguing as
there has been no direct and definitive observation of four-quark states
so far.%
\footnote{The four-quark state picture of the light scalar meson nonet
of ($\sigma$, $\kappa$, $a_0$, $f_0$) and exotic isotensor mesons were
discussed, e.g., in Refs.~\cite{BFSS99,APT05}.
A possible test for the structure of the light scalar mesons is
suggested in Ref.~\cite{CCY05}.}
In this picture, $D_s (2317)$ may be considered as a diquark-antidiquark
bound state as the diquark picture has been quite useful in describing
baryon spectroscopy, static properties, and decay mechanisms.
(See Ref.~\cite{APEFL:92} for a general review on the diquark.)

Indeed, Bracco {\it et al.\/}~\cite{BLMNN:05} recently performed the
QCD sum rule calculation of $D_s (2317)$ using the current
of scalar-diquark--scalar-antidiquark.
Their sum rules give the mass which agrees with the experimental value.
This suggests that the $D_s (2317)$ may be regarded as a four-quark state, 
especially a bound state of scalar-diquark and scalar-antidiquark.
However, there can be other possible currents for the diquark-antidiquark
system.
For example, if the light $[ud]$ diquark is isoscalar, it can have three
possible spin quantum numbers: scalar, pseudoscalar,
and vector~\cite{Griegel,SSV}.  
To form a four-quark state with $J^P=0^+$ in a simple approach,
therefore, one can combine two diquarks with such quantum numbers, which
yields three possible choices for the $D_s$ currents,  
namely, scalar-scalar, pseudoscalar-pseudoscalar, and vector-vector.
Since the diquarks in the four-quark state of our concern are isodoublet,
other choices like axial-vector--axial-vector combination may not be
excluded.
In the light quark sector, the axial-vector current gives isovector
diquark \cite{SSV}.

Of course, in the constituent quark picture the scalar-diquark may be favored
over the pseudoscalar diquark, because in the pseudoscalar channel the
upper component of one Dirac spinor is connected only with the lower
component of the other spinor.
Such combinations should vanish in the nonrelativistic limit.
However, QCD sum rules~\cite{QCDSR} deal with current quarks and the picture
based only on the constituent quarks should be carefully examined.
In particular, the standard nucleon current~\cite{Ioffe81} involves
a linear combination of two types of diquarks, scalar and pseudoscalar, and
the both contribute to the nucleon sum rule with a similar weight.
Therefore, a further test with the pseudoscalar-pseudoscalar current
is required for the $D_s$ sum rule.
The other four-quark currents with vector-vector and
axial-vector--axial-vector types are interesting as they cannot be ruled
out from the constituent quark picture.
Thus, dynamical calculations are needed to test these currents, which
would be important to understand the possible internal structure of $D_s$
meson when viewed as a tetra-quark state.
This will be eventually helpful to investigate the $D_s$ properties in
lattice QCD calculations~\cite{lattice}.

In this work, we test all the four currents for $D_s (2317)$ in QCD sum rules.  
For this purpose, we first improve the previous sum rule calculation of
Ref.~\cite{BLMNN:05} by including higher order terms in the operator product
expansion (OPE) that might be non-negligible for sum-rule predictions.
We then construct three more sum rules with pseudoscalar-pseudoscalar, 
vector-vector, and axial-vector--axial-vector currents.
By scrutinizing those sum rules, we hope to eliminate certain diquark
combinations as the main composition of the tetraquark $D_s$, which may give
a clue for the internal structure of $D_s$ meson.

\section{QCD sum rules for $\bm{D_s (2317)}$}

The quantum numbers of $D_s (2317)$ are believed to be $I(J^P) =
0(0^+)$~\cite{PDG04}.
The four-quark current with these quantum numbers can be constructed by
combining the diquark $[cu]$ and the antidiquark $[\bar{s}\bar{u}]$.
(For the QCD sum rule calculation for the normal $D_s (1968)$ meson, see,
e.g., Ref.~\cite{DP93}.)
The isoscalar diquark in the light ($u,d$) sector is restricted
to the three types: scalar, pseudoscalar, and vector~\cite{Griegel,SSV}.
Thus, its straightforward extension leads to the corresponding
three types for $[cu]$ and $[\bar{s}\bar{u}]$.
In addition, the axial-vector diquark may be another possibility.
By combining the diquark and antidiquark with the same type to make the
$J^P = 0^+$ state, we have four possible choices for the four-quark current,
\begin{eqnarray}
J_k &=& \frac {1} {\sqrt{2}}\left[ \epsilon_{ade}  \epsilon_{afg} c^T_d
\Gamma_k u_e  \bar{s}_f {\tilde \Gamma}_k \bar{u}^T_g  + (u \to d) \right ] ,
\label{current}
\end{eqnarray}
where $k=1,\dots, 4$ and
${\tilde \Gamma}_k \equiv \gamma^0 \Gamma_k^\dagger \gamma^0$.
The color indices are represented by the subscripts and they are chosen
so that the diquark in color space belongs to
$\overline{\bf 3}_c$ and the antidiquark to ${\bf 3}_c$.
The Dirac matrices between the quarks can be $\Gamma_1=C$ (pseudoscalar),
$\Gamma_2=C\gamma_5$ (scalar), $\Gamma_3=C\gamma_5 \gamma_\mu$ (vector), 
and $\Gamma_4 = C\gamma_\mu$ (axial-vector), where $C$ is the charge
conjugation operator.
For the vector and axial-vector cases, the two vector indices must
be contracted to form a scalar state.
Note that $C$, $C\gamma_5 \gamma_i$, and $C\gamma_0$ are off-diagonal
matrices and therefore they connect the upper and lower components of
participating Dirac spinors.
The other matrices, $C\gamma_5$, $C\gamma_5 \gamma_0$, and $C\gamma_i$ 
are diagonal so only the upper (or lower) components of the Dirac spinors
can be connected to each other.
Note that, to form an isoscalar four-quark current, the $(u\rightarrow
d)$ term is added in Eq.~(\ref{current}).

In constructing sum rules, we consider the following correlation
function,
\begin{eqnarray}
\Pi_k (p) = i \int d^4x\, e^{ip x} \langle 0 \mid T[ J_k (x) J^\dagger_k (0)]
\mid 0 \rangle  .
\end{eqnarray}
The time-ordering of quark fields is evaluated by the OPE.
This OPE is then matched with the hadronic expression of the
correlation function via the Borel-weighted sum rule,
\begin{eqnarray}
\int^{S_0}_{m_c^2} dp^2 e^{-p^2/M^2} {1\over \pi}
\,\mbox{Im} \left[ \Pi^{\rm phen} (p^2) - \Pi_k^{\rm OPE} (p^2) \right] =0 ,
\end{eqnarray}
where $M$ denotes the Borel mass and $m_c$ the charm quark mass.
The phenomenological side $\Pi^{\rm phen} (p^2)$ contains the
contribution from the low-lying resonance of our concern as well as
higher resonances or multi-meson continuum states.
The contributions other than the low-lying resonance have been subtracted
according to the QCD duality assumption, which introduces the continuum
threshold $S_0$.
With this relation, we can extract the $D_s$ mass, $m_{D_s}^{}$, from
\begin{equation}
|\lambda|^2 e^{-m_{D_s}^2 /M^2} = \int^{S_0}_{m_c^2}
dp^2 e^{-p^2/M^2} {1\over \pi} \,\mbox{Im} \left[ \Pi_k^{\rm OPE} (p^2)
\right],
\label{sum}
\end{equation}
where $\lambda$ represents the coupling strength of the interpolating
field to the physical $D_s$ state.
Specifically, we take a derivative with respect to $1/M^2$ and divide the
resulting equation by Eq.~(\ref{sum}) to get the sum rule for $m_{D_s}^2$.
Depending on the interpolating fields characterized by the index $k$, we
can construct four separate sum rules.
To get a sensible result, one has to check that the right-hand side of
Eq.~(\ref{sum}) is positive as constrained by the left-hand side.

\begin{figure}[t]
\centering \epsfig{file=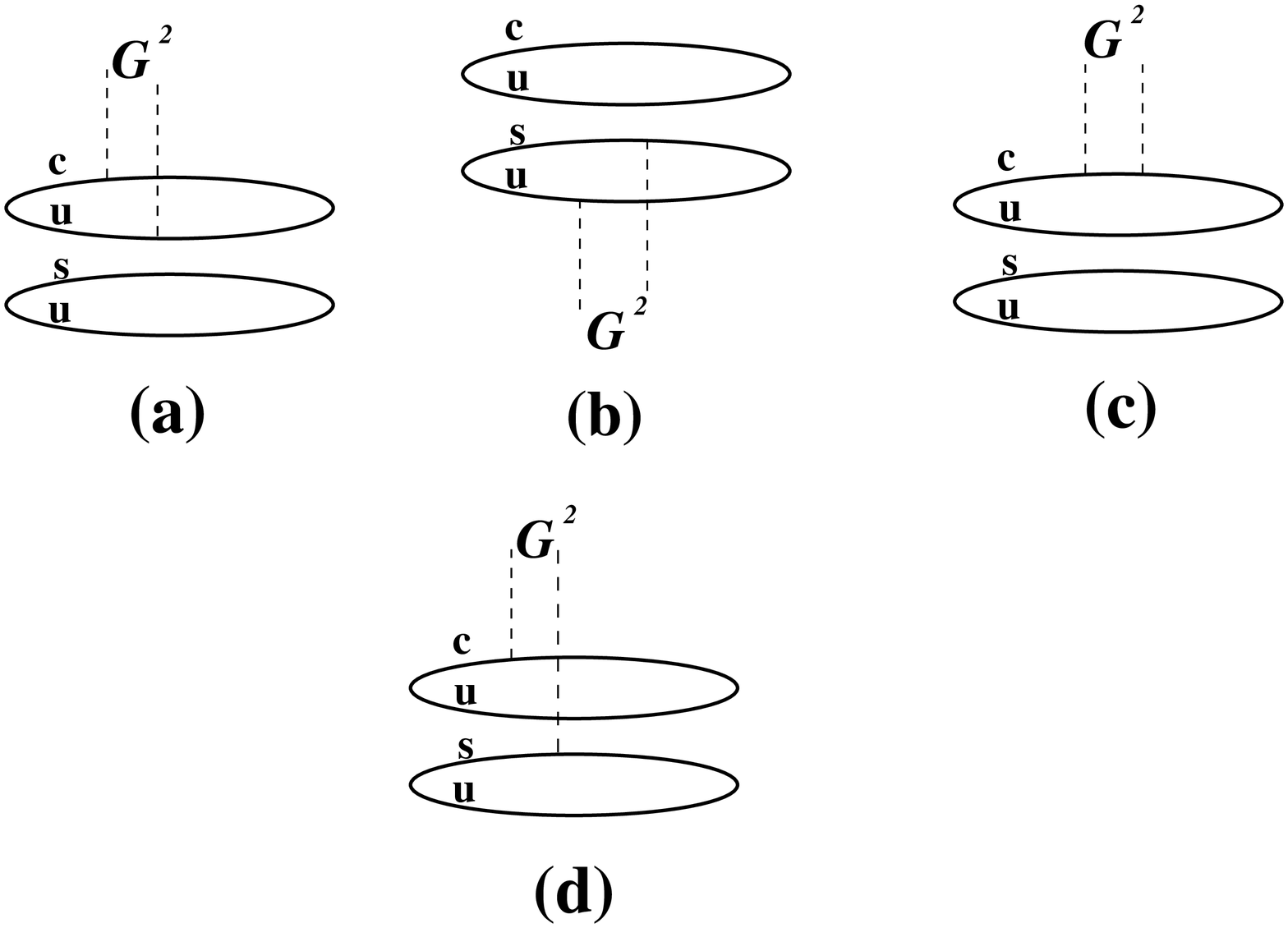, width=0.5\textwidth}
\caption{Diagrams that contribute to the gluon condensate.
The solid lines denote quark propagators with specified flavor and the
dashed lines are gluons.
When the gluon condensate is formed by taking one gluon from one diquark
and the other gluon from the other diquark, we have four possibilities.
(d) shows one of such possibilities.}
\label{fig1}
\end{figure}

The OPE calculation can be done straightforwardly using the same
technique developed in Ref.~\cite{KLO:04}.
Namely, we use the momentum-space expression for the charm-quark propagator
to keep the charm quark mass finite.
For the light-quark part, we calculate in the coordinate-space, which is
then Fourier-transformed to the momentum space in the $D$-dimension.
The resulting light-quark part is combined with the charm-quark part before
dimensionally regularized at $D=4$.
The OPE for the scalar-scalar correlator $\Pi_2^{\rm OPE}$ is obtained by
summing up the following terms,
\begin{widetext}
\begin{eqnarray}
&&\frac{1}{\pi} \,\mbox{Im}\Pi^{\rm pert}_2
=-\frac{1}{3\times 2^{10} \pi^6}
\int^{\Lambda}_0 du \left [ \frac{3}{2} u(1-u)p^2 -
\frac{1}{2} u m_c^2 \right ]
\left ( \frac {-L}{1-u}\right )^3,
\nonumber \\
&&\frac{1}{\pi} \,\mbox{Im}\Pi^{ m_s^2}_2
=
-\frac{m_s^2}{2^7 \pi^6}
\int^{\Lambda}_0 du \left [ -\frac{5}{6}u(1-u)p^2+
\frac{1}{3}u m_c^2 \right ] \left ( \frac {-L}{1-u}\right )^2,
\nonumber \\
&&\frac{1}{\pi} \,\mbox{Im}\Pi^{ \langle \bar {q} q \rangle}_2
=
\frac{m_c \langle \bar {q} q \rangle}{2^6 \pi^4}
\int^{\Lambda}_0 du \left ( \frac {-L}{1-u}\right )^2
-\frac{m_s \langle \bar {q} q \rangle}{2^5 \pi^4}
\int^{\Lambda}_0 du [2 u(1-u)p^2-u m_c^2 ] \frac {-L}{1-u},
\nonumber \\
&&\frac{1}{\pi} \,\mbox{Im}\Pi^{ \langle \bar {s} s \rangle}_2
=\frac{m_s \langle \bar {s} s \rangle}{2^6 \pi^4}
\int^{\Lambda}_0 du [2 u(1-u)p^2-u m_c^2 ] \frac {-L}{1-u},
\nonumber \\
&&\frac{1}{\pi} \,\mbox{Im}\Pi^{ \langle G^2 \rangle}_2 (a,b)
=-\frac{1}{2^9 \pi^4}
\left \langle \frac{\alpha_s}{\pi} G^2 \right \rangle
\int^{\Lambda}_0 du \left (1 -\frac{2u}{3}\right )
[2 u(1-u)p^2-u m_c^2 ]\frac {-L}{1-u},
\nonumber \\
&&\frac{1}{\pi} \,\mbox{Im}\Pi^{ \langle G^2 \rangle}_2 (c)
=-\frac{m_c^2}{2^{10} 3^2 \pi^4}
\left \langle \frac{\alpha_s}{\pi} G^2 \right \rangle
\int^{\Lambda}_0 du
[3 u(1-u)p^2-2 u m_c^2 ] \left( \frac {u}{1-u} \right )^3,
\nonumber \\
&&\frac{1}{\pi} \,\mbox{Im}\Pi^{ \langle G^2 \rangle}_2 (d)
= 0  ,
\nonumber \\
&&
\frac{1}{\pi} \,\mbox{Im}\Pi^{ \langle \bar {q} G q \rangle}_2
=  m_c \frac{\langle \bar {q} g_s \sigma  G q \rangle}{2^7 \pi^4}
\int^{\Lambda}_0 du \frac{u}{1-u}~\frac{-L}{1-u},
\nonumber \\
&&\frac{1}{\pi} \,\mbox{Im}\Pi^{ \langle \bar {q} D^2 q \rangle}_2
= m_c \frac{\langle \bar {q} D^2 q \rangle}{2^5 \pi^4}
\int^{\Lambda}_0 du \frac{-L}{1-u},
\nonumber \\
&&
\frac{1}{\pi} \,\mbox{Im}\Pi^{ \langle \bar {q} q \rangle^2}_2
=-\frac{\langle \bar {q} q \rangle \langle \bar {s} s \rangle}{12 \pi^2}
\int^{\Lambda}_0 du [-3 u(1-u)p^2+2 u m_c^2],
\nonumber \\
&&\frac{1}{\pi} \,\mbox{Im}\Pi^{ \langle \bar {q} q \rangle^3}_2
=-\frac{1}{9} m_c \langle \bar {q} q \rangle^2 \langle \bar {s} s \rangle
\delta(p^2-m_c^2),
\nonumber \\
&&\frac{1}{\pi} \,\mbox{Im}\Pi^{ \langle \bar {q} q \rangle
\langle \bar {q}G q \rangle}_2
=-\frac{1}{48\pi^2} \left [\langle \bar {s} s \rangle
\langle \bar {q}g_s \sigma  G q \rangle + \langle \bar {q} q \rangle
\langle \bar {s}g_s \sigma  G s \rangle \right ]
\left [\frac{m_c^4}{p^4} +1 \right ] .
\end{eqnarray}
\end{widetext}
Here $L\equiv u(1-u)p^2 - u m_c^2$ and $\Lambda\equiv
1-m_c^2/p^2$.
Compared with the OPE of Ref.~\cite{BLMNN:05}, the terms containing
$m_s^2$ and $m_s\langle {\bar s}s \rangle$ are new, but we found that their
contribution is not important.
In addition, we have higher order OPE containing quark condensate like
$\langle {\bar q} q \rangle^3$ and $\langle {\bar q} q \rangle \langle
{\bar q} G q \rangle$, which can constitute nontrivial contributions in
stabilizing the Borel curves.
Using the relation $\langle \bar{q} D^2 q \rangle =\langle \bar{q} g_s
\sigma G q \rangle/2$, one can combine the two terms at dimension 5.
We have written these terms separately because they are differently related
to the corresponding terms in the vector and axial-vector channels. 
Figure~\ref{fig1} shows the diagrams that contribute to the gluon condensate.
In the equations above, we separate the gluon contribution into two terms,
one for Figs.~\ref{fig1}(a) and (b), and the other for Fig.~\ref{fig1}(c).
Note that the contribution from Fig.~\ref{fig1}(d) vanishes in the
scalar-scalar channel but it is non-zero in the vector-vector channel.
We did not calculate the diagrams where two gluons emitted from a light
quark propagator as they should be parts of the quark condensate.

For the correlators with the other currents, $\Pi_1$ (pseudoscalar),
$\Pi_3$ (vector), and $\Pi_4$ (axial-vector), most of the OPE can be
obtained from $\Pi_2$ by careful inspection.
For the pseudoscalar correlator, we find
\begin{eqnarray}
&&\Pi_1^{\rm pert} = \Pi_2^{\rm pert} ,
\quad
\Pi_1^{m_s^2} = \Pi_2^{m_s^2} ,
\quad
\Pi_1^{\langle \bar{q} q\rangle} =
-\Pi_2^{\langle \bar{q} q\rangle} ,
\nonumber \\ &&
\Pi_1^{\langle \bar{s} s\rangle} =
\Pi_2^{\langle \bar{s} s\rangle} ,
\quad
\Pi_1^{\langle G^2 \rangle} (a,b) =
\Pi_2^{\langle G^2 \rangle} (a,b) ,
\nonumber \\ &&
\Pi_1^{\langle G^2 \rangle} (c) =
\Pi_2^{\langle G^2 \rangle} (c) ,
\quad
\Pi_1^{\langle \bar{q} D^2 q \rangle} =
-\Pi_2^{\langle \bar{q} D^2 q \rangle} ,
\nonumber \\ &&
\Pi_1^{\langle \bar{q} G q \rangle} =
-\Pi_2^{\langle \bar{q} G q \rangle} ,
\quad
\Pi_1^{\langle \bar{q} q \rangle^2} =
-\Pi_2^{\langle \bar{q} q \rangle^2} ,
\nonumber \\ &&
\Pi_1^{\langle \bar{q} q \rangle \langle \bar{q} G q \rangle} =
-\Pi_2^{\langle \bar{q} q \rangle \langle \bar{q} G q \rangle} ,
\quad
\Pi_1^{\langle \bar{q} q \rangle^3} =
\Pi_2^{\langle \bar{q} q \rangle^3} ,
\label{sign}
\end{eqnarray}
while we obtain for vector and axial-vector correlators
\begin{eqnarray}
&&\Pi_3^{\rm pert} = \Pi_4^{\rm pert}=4 \Pi_2^{\rm pert} ,
\quad
\Pi_3^{m_s^2}=\Pi_4^{m_s^2} = 4 \Pi_2^{m_s^2} ,
\nonumber \\
&&\Pi_3^{\langle \bar{q} q\rangle} = -\Pi_4^{\langle \bar{q} q\rangle} =
2 \Pi_2^{\langle \bar{q} q\rangle} ,
\quad
\Pi_3^{\langle \bar{s} s\rangle} = \Pi_4^{\langle \bar{s} s\rangle} =
4 \Pi_2^{\langle \bar{s} s\rangle}  ,
\nonumber \\
&&\Pi_3^{\langle G^2 \rangle} (a,b) = \Pi_4^{\langle G^2 \rangle} (a,b) =0 ,
\nonumber \\
&&\Pi_3^{\langle G^2 \rangle} (c,d) =\Pi_4^{\langle G^2 \rangle} (c,d)=
2 \Pi_2^{\langle G^2 \rangle}(a,b)+4 \Pi_2^{\langle G^2 \rangle}(c) ,
\nonumber \\
&& \Pi_3^{\langle \bar{q} D^2 q \rangle}
= -\Pi_4^{\langle \bar{q} D^2 q \rangle}
= 2 \Pi_2^{\langle \bar{q} D^2 q \rangle}  ,
\nonumber \\
&&\Pi_3^{\langle \bar{q} G q \rangle} =-\Pi_4^{\langle \bar{q} G q \rangle}
= -\Pi_2^{\langle \bar{q} D^2 q \rangle}  ,
\nonumber \\
&&\Pi_3^{\langle \bar{q} q \rangle^2} = -\Pi_4^{\langle \bar{q} q \rangle^2}
=
2 \Pi_2^{\langle \bar{q} q \rangle^2}  ,
\nonumber \\
&&\Pi_3^{\langle \bar{q} q \rangle^3} = \Pi_4^{\langle \bar{q} q \rangle^3}
= 4 \Pi_2^{\langle \bar{q} q \rangle^3} . 
\label{sign2}
\end{eqnarray}
In the above equations, $\Pi_{3,4}^{\langle G^2 \rangle}$ and 
$\Pi_{3,4}^{\langle \bar{q} G q \rangle}$ are calculated directly and their
relations to those of the scalar OPE look different from the others.
In addition, the $\langle \bar{q} q \rangle \langle \bar{q} G q \rangle$
term in the vector and axial-vector correlator is not simply related
to that of the scalar correlator.
Explicitly, it reads
\begin{eqnarray}
&&\frac{1}{\pi} \,\mbox{Im}\Pi^{ \langle \bar {q} q \rangle
\langle \bar {q}G q \rangle}_3 
= -\frac{1}{\pi} \,\mbox{Im}\Pi^{ \langle \bar {q} q \rangle
\langle \bar {q}G q \rangle}_4
\nonumber \\
&&
=\frac{1}{96\pi^2} \left[\langle \bar {s} s \rangle
\langle \bar {q}g_s \sigma  G q \rangle + \langle \bar {q} q \rangle
\langle \bar {s}g_s \sigma  G s \rangle \right]
\left[\frac{5m_c^4}{p^4} -\frac{3m_c^2}{p^2}+4 \right] .
\end{eqnarray}

\section{Results and discussion}

\begin{figure}[t]
\centering \epsfig{file=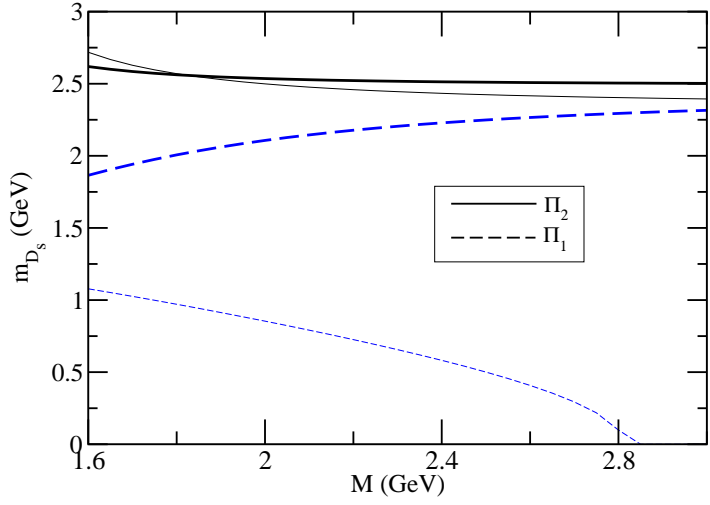, width=0.5\textwidth, angle=-90}
\caption{(Color online)
The Borel curves for the $D_s$ mass in the scalar and
pseudoscalar channels.
The solid curves are for the $\Pi_2$ (scalar) sum rule.
The dashed curves are for the $\Pi_1$ (pseudoscalar) sum rule.
In each case, the thick line is for $\sqrt{S_0}=2.7$ GeV and the thin line is
for $\sqrt{S_0}=2.4$ GeV.
We observe that for the $\Pi_1$ sum rule the extracted
mass is highly sensitive to the continuum threshold.
But for the $\Pi_2$ sum rule the Borel curves are stable and not so sensitive
to the continuum threshold.}
\label{fig2}
\end{figure}

Given in Fig.~\ref{fig2} are the Borel curves for the mass $m_{D_s}^{}$
in the scalar-scalar and pseudoscalar-pseudoscalar cases.
They are obtained by using the following standard values for the
QCD parameters as used in Ref.~\cite{KLO:04},
\begin{eqnarray}
&&\left \langle {\alpha_s \over \pi} G^2
\right \rangle = (0.33 \mbox{ GeV})^4  ,
\nonumber \\
&&m_c = 1.26 \mbox{ GeV} , \qquad m_s=0.15 \mbox{ GeV},
\nonumber \\
&& \langle \bar {q} q \rangle = -(0.23\mbox{ GeV})^3 ,
\qquad
\langle \bar {s} s \rangle = 0.8 \,\langle \bar {q} q \rangle,
\quad
\nonumber \\
&& \langle \bar {q}g_s \sigma G q \rangle
= m_0^2 \langle \bar {q} q \rangle ,
\qquad
\langle \bar {s}g_s \sigma G s \rangle
=0.8 \, \langle \bar {q}g_s \sigma G q \rangle\ .
\label{eq:param}
\end{eqnarray}
The parameter $m_0^2$ denotes the quark virtuality which
is normally taken to be $0.8$ GeV$^2$~\cite{JCFG:92}.
To show the sensitivity to the continuum threshold, we vary
$\sqrt{S_0}$ from 2.4 GeV (thin lines) to a somewhat larger value,
2.7 GeV (thick lines).
The lower bound, 2.4 GeV, is slightly above the $KD$ threshold.
The $\Pi_2$ sum rule shown by the two solid lines clearly exhibits the
Borel stability yielding $m_{D_s}^{}$ around 2.4--2.5 GeV depending on the
continuum.
The two solid lines (thin and thick lines) are quite close to each other
indicating that this sum rule is insensitive to the continuum threshold.
Our result is not so sensitive to the uncertainties of the most QCD
parameters given above except for the parameter associated with the
quark-gluon mixed condensate, $m_0^2$.
If somewhat lower value $m_0^2=0.6$ GeV$^2$ is used in this sum rule,
the extracted mass with $\sqrt{S_0}=2.4$ GeV is found to be around
2.27 GeV.
In fact, the quark-gluon mixed condensate plays a crucial role in making
the right-hand side of Eq.~(\ref{sum}) be positive as constrained by
the left-hand side.
We found that, without the quark-gluon condensate, this constraint
is not satisfied in this particular sum rule.
Our stable result may imply that the scalar-scalar current couples strongly
to the low-lying pole while its couplings to higher resonances or continuum
states are suppressed.
Therefore, the scalar-scalar current might be an optimal candidate for the
$J^P = 0^+$ $D_s$ resonance.

\begin{figure}[t]
\centering \epsfig{file=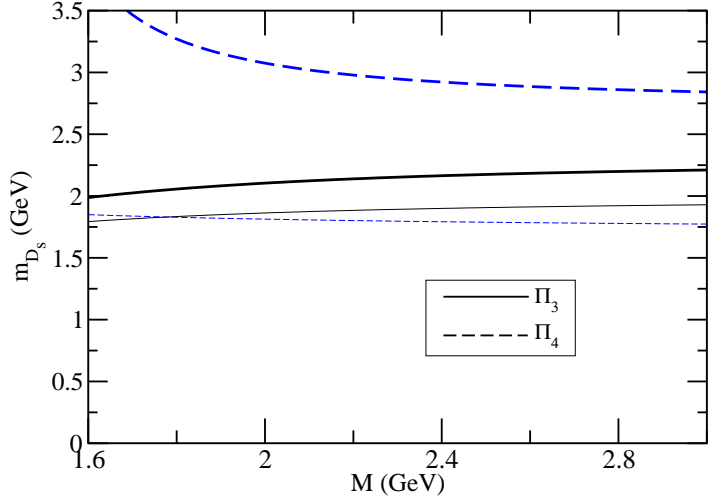, width=0.5\textwidth, angle=-90}
\caption{(Color online)
The Borel curves for the $D_s$ mass in the vector and
axial-vector channels.
The solid curves are for the $\Pi_3$ (vector) sum rule and
the dashed curves are for the $\Pi_4$ (axial-vector) sum rule.
The thick lines are for $\sqrt{S_0}=2.7$ GeV and the thin lines
for $\sqrt{S_0}=2.4$ GeV.}
\label{fig3}
\end{figure}

On the other hand, the Borel curves for the $\Pi_1$ sum rule given by
the dashed lines in Fig.~\ref{fig2} show strong sensitivity to the continuum
threshold.
The thick dashed line ($\sqrt{S_0} = 2.7$ GeV) is substantially different
from the thin dashed line ($\sqrt{S_0} = 2.4$ GeV).
Also, the strong variation of the curves with respect to the Borel mass
clearly shows the Borel instability.
{}From this sum rule, therefore, it is hard to extract any
stable value for the $D_s$ mass.  
In this case, the important contribution coming from the quark-gluon mixed
condensate changes its sign from that of $\Pi_2$ as can be seen from
Eq.~(\ref{sign}), which makes the Borel curves unstable.
Thus, the quark-gluon mixed condensate contributes differently in the
pseudoscalar sum rule.
Note that, when $\sqrt{S_0} = 2.4$ GeV, there is no extracted mass for
$M \ge 2.82$ GeV as it becomes imaginary. 
One way to understand the success (failure) of the scalar (pseudoscalar) 
sum rule may be the dominance of the nonrelativistic configuration of the
current for the $D_s$ sum rules.
Namely, the scalar current survives in the nonrelativistic limit while the
pseudoscalar current does not.

Figure~\ref{fig3} shows the results from the $\Pi_3$ and $\Pi_4$ sum rules. 
We observe that for the $\Pi_3$ sum rule the extracted mass is slightly
sensitive to the continuum threshold.
As we change $\sqrt{S_0}$ from 2.4 GeV to 2.7 GeV, the extracted mass is
in the range of 1.92 to 2.2 GeV, which is somewhat smaller than the
experimental $D_s$ mass and the value obtained with the $\Pi_2$ sum
rule.
But the extracted mass can be larger when a slightly larger continuum
threshold is used.
Furthermore, the Borel curves are quite stable with respect to the 
Borel mass and thus it can be another possible choice for the $D_s$ current.
In the $\Pi_4$ sum rule, however, the Borel curves are quite sensitive
to the continuum threshold as one can see from the dashed lines in
Fig.~\ref{fig3}.
The result from the $\Pi_4$ sum rule with $\sqrt{S_0}=2.4$ GeV is in fact
unphysical since the OPE is found to violate the positivity constraint
provided by Eq.~(\ref{sum}).
Thus, the current with axial-vector--axial-vector composition is ruled
out from the possible choices for the $D_s$ current.
If this result should be related to the dominance of the nonrelativistic
configuration as conjectured in the scalar and pseudoscalar channels,
then our results in the vector and axial-vector channels
can be understood from the dominance of the time component
of diquarks over their spatial components.

\begin{figure}[t]
\centering \epsfig{file=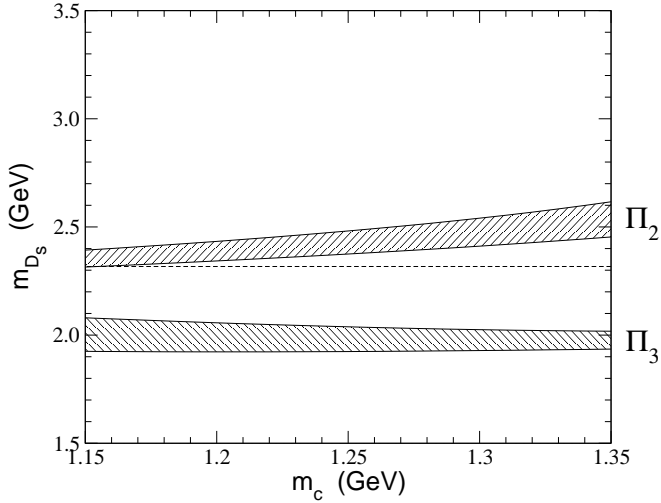, width=0.5\textwidth, angle=-90}
\caption{
The $D_s$ mass as a function of the charm quark mass $m_c$ with the
scalar-scalar ($\Pi_2$) current and with the vector-vector ($\Pi_3$)
current at the Borel mass $M=2.4$ GeV.
The shaded area for $\Pi_2$ current is obtained by varying $m_0^2$ from
0.7 GeV$^2$ to 0.9 GeV$^2$, while that for $\Pi_3$ current by varying
$\langle \bar{q} q \rangle = - (0.23 \mbox{ GeV})^3 \sim
- (0.25 \mbox{ GeV})^3$.
Other parameters are fixed as given in Eq.~(\ref{eq:param}) in each case
with $\sqrt{S_0} = 2.5$ GeV.
The dashed line denotes the experimental mass of $D_s(2317)$.}
\label{fig4}
\end{figure}

As we have discussed above, the obtained masses are mostly dependent of
the continuum threshold $S_0$.
In order to estimate the uncertainties arising from the other parameters
we plot the obtained $D_s$ mass as a function of the charm quark mass
$m_c$ in Fig.~\ref{fig4}.
Among the parameters given in Eq.~(\ref{eq:param}), we found that
the $D_s$ mass with $\Pi_2$ current is most sensitive to the quark
virtuality $m_0^2$, while the mass obtained with $\Pi_3$ is to the quark
condensate $\langle \bar{q} q \rangle$.
The shaded area in Fig.~\ref{fig4} are, therefore, obtained with $m_0^2
= 0.7 \mbox{ GeV}^2 \sim 0.9 \mbox{ GeV}^2$ for $\Pi_2$ and with
$\langle \bar{q} q \rangle = - (0.23 \mbox{ GeV})^3 \sim
- (0.25 \mbox{ GeV})^3$ for $\Pi_3$.
As the results are much less sensitive to the other parameters, they are
fixed as given in Eq.~(\ref{eq:param}) and
$\sqrt{S_0} = 2.5$ GeV is used.
Within the range of the parameters considered in this calculation,
Fig.~\ref{fig4} shows that the uncertainties are not large and actually
they are at the order of 5\% for a given continuum threshold $\sqrt{S_0}$.

In summary, we have performed QCD sum rule calculations for $D_s(2317)$
using four different tetra-quark currents with the final quantum numbers
$J^P = 0^+$.
The sum rule using the scalar-scalar current gives the stable Borel curves
with least sensitivity to the continuum threshold.
This slightly improves the previous calculation by Bracco
{\it et al.\/}~\cite{BLMNN:05} and supports the picture of $D_s(2317)$
as a bound state of scalar-diquark and scalar-antidiquark.
Also the vector-vector sum rule was found to give a stable result for the
$D_s$ mass with slightly lower value.
This implies that the vector-vector combination could be another
possible choice and cannot be simply ruled out for the $D_s$ current. 
Of course, the possibility that the physical $D_s$ state may be a
mixture of the two configurations cannot be excluded.
The other two sum rules using the pseudoscalar-pseudoscalar and
axial-vector--axial-vector currents yield unstable results with strong
sensitivity to the continuum threshold.
Therefore, they are unfavored as the main component of the $D_s(2317)$ meson.
Although we could not rule out the other models, such as quark-antiquark
description,%
\footnote{
The different expectations from the two-quark and four-quark picture of
$D_s(2317)$ in the $B$ meson decays were discussed in Ref.~\cite{CCY05}.
}
for the $D_s$ meson by this study, our results indicate the
possible four-quark structure with the scalar-scalar and/or vector-vector
diquarks of the $D_s$ meson.

\acknowledgments

Y.O. was supported by Forschungszentrum-J{\"u}lich, contract
No. 41445282 (COSY-058).


\end{document}